\documentclass[runningheads,a4paper]{llncs}

\usepackage{amssymb}
\usepackage{graphicx}

\usepackage{url}

\newcommand{\keywords}[1]{\par\addvspace\baselineskip
\noindent\keywordname\enspace\ignorespaces#1}

\begin{document}

\mainmatter  

\title{Bots, elections, and social media:\\a brief overview}

\titlerunning{}

\toctitle{Bots, elections, and social media: a brief overview}
\tocauthor{Emilio Ferrara}
\author{Emilio Ferrara%
}
\authorrunning{}

\institute{USC Information Sciences Institute\\
4676 Admiralty way, 1001, Marina del Rey, CA 90292, USA\\
\url{emiliofe@usc.edu}}

%
%

\maketitle

\begin{abstract}
Bots, software-controlled accounts that operate on social media, have been used to manipulate and deceive. We studied the characteristics and activity of bots around major political events, including elections in various countries. In this chapter, we summarize our findings of bot operations in the context of the 2016 and 2018 US Presidential and Midterm elections and the 2017 French Presidential election.
\\
\end{abstract}

\section{Introduction}

Social media have been widely portrayed as enablers of democracy~\cite{lazer2009life,kwak2010twitter,cha2010measuring,boyd2012critical,kumpel2015news}. 
In countries were freedom to communicate and organize lacked, social media provided a platform to openly discuss political~\cite{adamic2005political,effing2011social,bekafigo2013tweets,carlisle2013social,digrazia2013more,lutz2014beyond,yang2016social} and social issues~\cite{gonzalez2011dynamics,gonzalez2013broadcasters,conover2013geospatial,conover2013digital,varol2014evolution,barbera2015critical,theocharis2015using}, without fears for safety or retaliation. Such platforms have also been used to respond to crises and emergencies~\cite{sutton2008backchannels,yates2011emergency,gao2011harnessing,yin2012using,latonero2013emergency}. 
It is hard to overstate the importance of these platforms for the billions of people who use them every day, all over the world. 

However, as it happens with most powerful emerging technologies, the rise of popularity led to abuse. 
Concerns about the possibility of manipulating public opinion using social media have been brought a decade before they materialized~\cite{howard2006new}. 
Ample evidence was provided by the scientific community that social media can influence people's behaviors~\cite{aral2011creating,centola2011experimental,kramer2014experimental,ferrara2015measuring,monsted2017evidence,ferrara2016detection}.
These concerns have been corroborated by numerous recent studies~\cite{ratkiewicz2011truthy,metaxas2012social,el2013social,ferrara2015manipulation,howard2016bots,shorey2016automation,Varol2017,ferrara2017disinformation}. 

Social media can be used to reach millions of people using targeted strategies aimed to maximize the spread of a message. If the goal is to manipulate public opinion, one way to achieve it is by means of bots, software-controlled social media accounts whose goal is to mimic the characteristics of human users, while operating at much higher pace at substantially no downside for their operators. Bots can emulate all  basic human activity on social media platforms, and they become increasingly more sophisticated as new advancements in Artificial Intelligence emerge~\cite{hwang2012socialbots,messias2013you,ferrara2016rise,ICWSM1715587,stella2018bots}.

In this chapter, we focus on the use of bots to manipulate the political discourse.
The first anecdotal accounts of attempts to steer public opinion on Twitter date back to the 2010 US Midterm election~\cite{ratkiewicz2011detecting} and similarly during the 2010 US Senate special election in Massachusetts~\cite{mustafaraj2010obscurity,metaxas2012social}, where bots were used to generate artificial support for some candidates and to smear their opponents.

Attribution, \textit{i.e.,} the determination of the actors behind such operations, has proven challenging in most such cases~\cite{ferrara2016rise}. 
One notorious exception is represented by the attribution of an interference campaign occurred during the 2016 US Presidential election to a Russian-sponsored operation. This was as a result of a thorough investigation on Russian interference led by the US Senate Select Committee on Intelligence (SSCI). They found that ``The Russian government interfered in the 2016 U.S. presidential election with the goal of harming the campaign of Hillary Clinton, boosting the candidacy of Donald Trump, and increasing political and social discord in the United States.''\footnote{See Wikipedia: \url{https://en.wikipedia.org/wiki/Russian_interference_in_the_2016_United_States_elections}}
Numerous studies have investigated the events associated with this operation~\cite{kollanyi2016bots,bessi2016social,badawy2019falls}.

It is worth noting that bots have been used for other purposes, for example social spam and phishing~\cite{jagatic2007social,thomas2011suspended,song2011spam,jin2011data,yang2012analyzing,mukherjee2012spotting,thomas2013trafficking,ferrara2019history}. Albeit much work has been devoted to the challenges of detecting  social spam~\cite{markines2009social,gao2010detecting,zhang2012detecting} and spam bots~\cite{lee2010social,lee2010uncovering,stringhini2010detecting,boshmaf2011socialbot,mukherjee2012spotting}, only recently the research community started to investigate the effects that bots have on society, political discourse, and democracy. The goal of this chapter is to summarize some of the most important results in this space.

\subsection*{Contributions of this chapter}
The aim of this chapter is to connect results of our investigations into three major political events: \textit{(i)} the 2016 US Presidential election; \textit{(ii)} the 2017 French Presidential election; and \textit{(iii)} the 2018 US Midterm elections. We will discuss the role of bots in these events, and highlight the influence they had on the online political discourse. The contributions of this chapter are as follows:

\begin{itemize}
\item We first provide a brief overview of how bots operate and what are the challenges in detecting them. Several recent surveys have been published on the problem of characterizing and detecting bots \cite{stieglitz2017social,yang2019arming}, including our own on \textit{Communications of the ACM}~\cite{ferrara2016rise}.

\item We then illustrate our first, and maybe the most prominent, use case of bots-driven interference in political discourse, discussing how bots have been used during the 2016 US Presidential election to manipulate the discussion of the presidential candidates. This overview is based on our results that appeared prior to the November 8, 2016 election events~\cite{bessi2016social}.

\item We then illustrate how bots have been used to spread disinformation prior to the 2017 French Presidential election to smear Macron's public image. 

\item Finally, we overview recent results that suggest how bots have been evolving over the course of the last few years, focusing on the 2018 US Midterm elections, and we discuss the challenges associated to their detection.
\end{itemize}

\section{Anatomy of a bot}

\subsection{What is a bot}
In this chapter, we define as \textit{bot} (short for \textit{robot}, a.k.a., social bot, social media bot, social spam bot, or sybil account) a social media account that is predominantly controlled by software rather than a human user. Although the definition above inherently states nothing about the intents behind creating and operating a bot, according to published literature, malicious applications of bots are reported significantly more frequently than legitimate usage~\cite{ferrara2016rise,stieglitz2017social}. 

While in this chapter we will focus exclusively on bots that aim to manipulate the public discourse, it is worth nothing that some researchers have used bots for social good~\cite{monsted2017evidence,allem2017cigarette}, as illustrated by a recent taxonomy that explores the interplay between intent and characteristics of bots \cite{stieglitz2017social}.
Next, we describe some techniques to create and detect bots.

\subsection{How to create a bot}
In the early days of online social media, in the late 2000s, creating a bot was not a simple task: a skilled programmer would need to sift through various platforms' documentation to create a software capable of automatically interfacing with the front-end or the back-end, and operate functions in a human-like manner. 

These days, the landscape has completely changed: indeed, it has become increasingly simpler to deploy  bots, so that, in some cases, no coding skills are required to setup accounts that perform simple automated activities: tech blogs often post tutorials and ready-to-go tools for this purposes. Various source codes for sophisticated social media bots can be found online as well, ready to be customized and optimized by the more technically-savvy users~\cite{kollanyi2016bots}. 

We recently inspected same of the readily-available Twitter bot-making tools and compiled a non-comprehensive list of capabilities they provide \cite{bessi2016social,ferrara2017disinformation}.

Most of these bots can run within cloud services or infrastructures like \textit{Amazon Web Services} (AWS) or Heroku, making it more difficult to block them when they violate the Terms of Service of the platform where they are deployed. 

A very recent trend is that of providing Bot-As-A-Service (BaaS): Advanced conversational bots powered by sophisticated Artificial Intelligence are provided by companies like \textit{ChatBots.io} that can be used to carry digital spam campaigns \cite{ferrara2019history} and scale such operations by automatically engaging with online users. 

Finally, the increasing sophistication of Artificial Intelligence (AI) models, in particular in the area of \textit{neural-based natural language generation}, and the availability of large pre-trained models such as OpenAI's GPT-2 \cite{radford2019language}, makes it easy to programmatically generate text content. This can be used to program bots that produce genuine-looking short texts on platforms like Twitter, making it harder to distinguish between human and automated accounts~\cite{alarifi2016twitter}.

\subsection{How to detect bots}
The detection of bots in online social media platform has proven a challenging task. For this reason, it has attracted a lot of attention from the computing research community. Even DARPA, the U.S. \textit{Defense Advanced Research Projects Agency}, became interested and organized the 2016 DARPA Twitter Bot Detection~\cite{subrahmanian2016darpa}, with University of Maryland, University of Southern California, and Indiana University topping the challenge,  focused on detecting bots pushing anti vaccination campaigns. Large botnets have been identified on Twitter, from dormant  \cite{echeverria2017discovery,echeverria2017discovery}, to very active \cite{abokhodair2015dissecting}.

The literature on bot detection has become very extensive. We tried to summarize the most relevant approaches in a survey paper recently appeared on the \textit{Communications of the ACM}~\cite{ferrara2016rise}: In that review, we proposed a simple taxonomy to divide the bot detection approaches into three classes: \emph{(i)} bot detection systems based on social network information; \emph{(ii)} systems based on crowd-sourcing and leveraging human intelligence; \emph{(iii)} machine learning methods based on the identification of highly-predictive features that discriminate between bots and humans. We refer the interested reader to that review for a deeper analysis of this problem \cite{ferrara2016rise}.
Other  recent surveys propose complementary or alternative taxonomies that are worth considering as well~\cite{stieglitz2017social,cresci2017paradigm,cresci2017paradigm,yang2019arming}.

As of today, there are a few publicly-available tools that allow to do bot detection and study social media manipulation, including \textit{(i)} Botometer,\footnote{Botometer: \url{https://botometer.iuni.iu.edu/}}  a popular bot detection tool developed at Indiana University \cite{davis2016botornot}, \textit{(ii)} BotSlayer,\footnote{BotSlayer: \url{https://osome.iuni.iu.edu/tools/botslayer/}} an application that helps track and detect potential manipulation of information spreading on Twitter, and \textit{(iii)} the Bot Repository,\footnote{Bot Repository: \url{https://botometer.iuni.iu.edu/bot-repository/}} a centralized database to share annotated datasets of Twitter social bots. 

In conclusion, several algorithms have been published to detect bots using sophisticated machine learning techniques including deep learning~\cite{kudugunta2018deep}, anomaly detection~\cite{minnich2017botwalk,gilani2017classification,de2018lobo}, and time series analysis~\cite{chavoshi2016debot,stukal2017detecting}.

\section{Social media manipulation}

Bots have been reportedly used  to interfere in political discussions online, for example by creating the impression of an organic support behind certain political actors~\cite{mustafaraj2010obscurity,ratkiewicz2011detecting,ratkiewicz2011truthy,metaxas2012social}. However, the apparent support can be artificially generated by means of orchestrated campaigns with the help of bots. This strategy is commonly referred to as social media \textit{astroturf}~\cite{ratkiewicz2011truthy}. 

\subsection{2016 US Presidential Election}
Our analysis of social media campaigns during the 2016 US Presidential Election revealed the presence of social bots. We here summarize our findings first published in \cite{bessi2016social}, discussing data collection, bot detection, and sentiment analysis.

\subsubsection*{Data Collection.} 
We manually crafted a list of hashtags and keywords related to the 2016 US Presidential Election with 23 terms in total, including 5 terms specifically for the Republican Party nominee Donald Trump, 4 terms for the Democratic Party nominee Hillary Clinton, and the remainder terms relative to the four presidential debates. The complete list of search terms is reported in our paper~\cite{bessi2016social}. 
By querying the Twitter Search API between September 16 and October 21, 2016, we collected a large dataset. After post-processing and cleaning procedures, we studied a corpus constituted by 20.7 million tweets posted by nearly 2.8 million distinct users. 

\subsubsection*{Bot detection.}
We used Botometer \textit{v1} (the version available in 2016) to determine the likelihoood that the most active accounts in this dataset were controlled by humans or were otherwise bots. To label accounts as bots, we use the fifty-percent threshold---which has proven effective in prior studies~\cite{ferrara2016rise,davis2016botornot}---an account was considered to be a bot if the bot score was above 0.5. 
Due to the Twitter API limitations, it would have been impossible to test all the 2.78 million accounts in short time. Therefore, we tested the top 50 thousand accounts ranked by activity volume, which account for roughly 2\% of the entire population and yet are responsible for producing over 12.6 million tweets, which is about 60\% of the total conversation. Of the top 50 thousand accounts, Botometer classified as likely bots a total of 7,183 users (nearly 15\%), responsible for 2,330,252 tweets; 2,654 users were classified as undecided, because their scores did not significantly diverge from the classification threshold of 0.5; the rest---about 40 thousand users (responsible for just 10.3 million tweets, less than 50\% of the total)---were labeled as humans. Additional statistics are summarized in our paper~\cite{bessi2016social}.

\subsubsection*{Sentiment analysis.}
We leveraged sentiment analysis to quantify how bots (resp., humans) discussed the candidates. We used SentiStrength~\cite{thelwall2010sentiment} to derive the sentiment scores of each tweet in our dataset. This toolkit is especially optimized to infer sentiment in short informal texts, thus ideally suited for social media.  We tested it extensively in prior studies on the effect of sentiment on tweets' diffusion~\cite{ferrara2015measuring,ferrara2015quantifying}.
The algorithm assigns to each tweet $t$ a positive $P^+(t)$ and negative $P^-(t)$ polarity score, both ranging between 1 (neutral) and 5 (strongly positive/negative). Starting from the polarity scores, we captured the emotional dimension of each tweet $t$ with one single measure, the sentiment score $S(t)$, defined as the difference between positive and negative polarity scores: $S(t) = P^+(t) - P^-(t)$.
The above-defined score ranges between -4 and +4. The negative extreme indicates a strongly negative tweet, and occurs when $P^+(t)=1$ and $P^-(t)=5$. Vice-versa, the positive extreme identifies a strongly positive tweet labeled with $P^+(t)=5$ and $P^-(t)=1$. In the case $P^+(t)=P^-(t)$---positive and negative sentiment scores for a tweet $t$ are the same---the sentiment $S(t)=0$ of tweet $t$ is considered neutral as the polarities cancel each other out.

\begin{figure*}[t]
\includegraphics[width=\columnwidth]{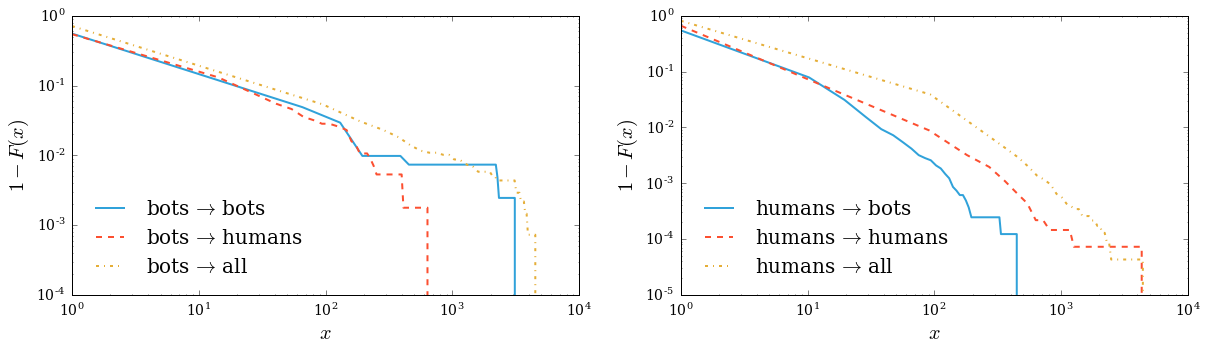}
\caption{Complementary cumulative distribution function (CCDF) of replies interactions generated by bots (left) and humans (right) (from~\cite{bessi2016social}).}
\label{fig:321}
\end{figure*}

\begin{figure*}[t]
\includegraphics[width=\columnwidth]{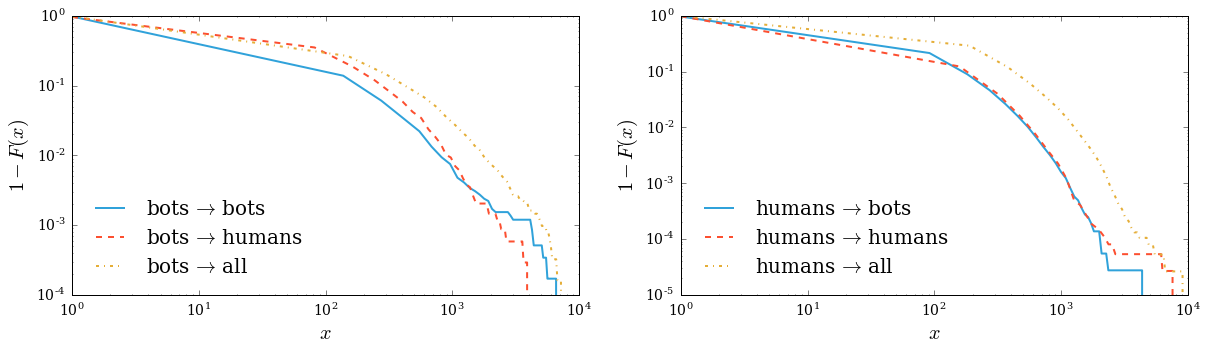}
\caption{Complementary cumulative distribution function (CCDF) of retweets interactions generated by bots (left) and humans (right) (from~\cite{bessi2016social}).}
\label{fig:322}
\end{figure*}

\subsubsection*{Partisanship and Supporting Activity.}
We used a simple heuristic based on the 5 Trump-supporting hashtags  and the 4 Clinton-supporting   to attribute user partisanships. For each user, we calculated their top 10 most used hashtags: If the majority  supported one particular candidate, we assigned the given user to that political group (Clinton or Trump supporter). Compared to network-based techniques~\cite{conover2011political,badawy2018analyzing}, this simple  partisanship assignment yielded a smaller yet higher-confidence annotated dataset,  constituted by 7,112 Clinton supporters (590 bots and 6,522 humans) and 17,202 Trump supporters (1,867 bots and 15,335 humans).

\subsubsection*{Summary of Results: Engagement.}

Figure~\ref{fig:321} and Figure~\ref{fig:322} illustrate the Complementary Cumulative Distribution Functions (CCDFs) of replies and retweets initiated by bots and humans in three categories: (i) within group (for example bot-bot, or human-human); (ii) across groups (e.g., bot-human, or human-bot); and, (iii) total (i.e., bot-all and human-all). The  heavy-tailed distributions, typically observed in social systems, appear in both. Hence, further inspection of Fig.~\ref{fig:321} suggests that \textit{(i)} humans  replied significantly more to other humans than to bots and, (ii) conversely, bots receive replies from other bots significantly more than from humans. One  hypothesis is that unsophisticated bots could not produce engaging-enough questions to foster meaningful exchanges with humans. 

Figure~\ref{fig:322}, however, demonstrates that retweets were a much more vulnerable mode of information diffusion: there is no statistically significant difference in the amount of retweets that humans generated by resharing content produced by other humans or by bots. In fact, humans and bots retweeted each other substantially at the same rate. This suggests that bots were very effective at getting their messages reshared in the human communication channels. 

Our study highlighted a vulnerability in the information ecosystem at that time, namely that content was reshared often without a thorough scrutiny on the information source. Several subsequent studies hypothesized that bots may have played a role in the spread of false news and unverified rumors~\cite{shao2018spread,vosoughi2018spread}.

\subsubsection*{Summary of Results: Sentiment.}

\begin{figure*}[t]
\includegraphics[width=\columnwidth]{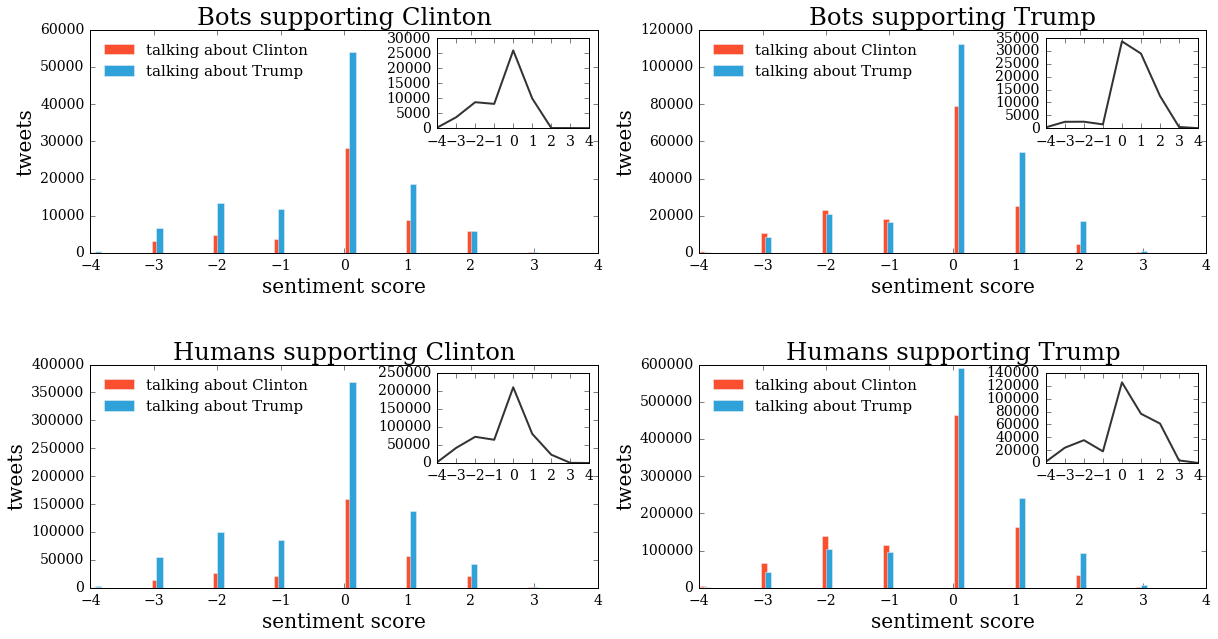}
\caption{Distributions of the sentiment of bots (top) and humans (bottom) supporting the two presidential candidates. The main histograms show the disaggregated volumes of tweets talking about the two candidates separately, while the insets show the absolute value of the difference between them (from~\cite{bessi2016social}).}
\label{fig:323}
\end{figure*}

We further explored how bots and humans talked about the two presidential candidates. Next, we show the sentiment analysis results based on \textit{SentiStrength}. Figure~\ref{fig:323} illustrates four settings: the top (resp., bottom) two panels show the sentiment of the tweets produced by the bots (resp., humans). Furthermore, the two left (resp., right) panels show the support for Clinton (resp., Trump). The main histograms in each panel show the volume of tweets about Clinton or Trump, separately, whereas the insets show the difference between the two. By contrasting the left and right panels we note that the tweets mentioning Trump are significantly more positive than those mentioning Clinton, regardless of whether the source is human or bot. However, bots tweeting about Trump generated almost no negative tweets and indeed produced the most positive set of tweets in the entire dataset  (about 200,000 or nearly two-third of the total). 

The fact that bots produce systematically more positive content in support of a candidate can bias the perception of the individuals exposed to it, suggesting that there exists an organic, grassroots support for a given candidate, while in reality it is in part artificially inflated. Our paper reports various examples of tweets generated by bots, and the candidate they support~\cite{bessi2016social}.

\subsection{2017 French Presidential Election}
A subsequent analysis of the Twitter ecosystem highlighted the presence and effects of bots prior to the 2017 French Presidential Election. We next report our findings summarizing the results published in 2017~\cite{ferrara2017disinformation}. We provide a characterization of both the bots and the users who engaged with them.

\subsubsection*{Data Collection.}
By following the same strategy as in the 2016 US Presidential election~\cite{bessi2016social}, we manually selected a set of hashtags and keywords related to the 2017 French Presidential Election. By construction, the list  contained a roughly equal number of terms associated with each of the two candidates, namely Marine Le Pen and Emmanuel Macron, and various general election-related terms: we ultimately identified 23 terms, listed in our paper~\cite{ferrara2017disinformation}. We  collected data by using the Twitter Search API, from April 27 to the end of election day, on  May 7, 2017: This procedure yielded a  dataset containing approximately 17 million unique tweets, posted by 2,068,728 million unique users. Part of this corpus is a subset of tweets associated with the \textit{MacronLeaks} disinformation campaign, whose details are described in our paper~\cite{ferrara2017disinformation}. The timeline of the volume of posted tweets is illustrated in Figure~\ref{fig:macron_timeline}.

\subsubsection*{Bot Detection.}
Due to the limitations of the Twitter API, and the time restrictions for this short period of unfolding events, we were unable to run in real time the bot detection relying upon Botometer. For this reason, we carried out a post-hoc bot detection on the dataset using an offline version of the bot-detection algorithm inspired by Botometer's rationale. Specifically, we  exclusively leveraged user metadata and activity features to create a simple yet effective bot detection classifier, trained on same data as Botometer, which is detailed in our paper~\cite{ferrara2017disinformation}. We validated its classification accuracy and assessed that it was similar to Botometer's performance, with above 80 percent in both accuracy and AUC-ROC scores. Manual validation corroborated the performance analysis. 
Hence, we used this simplified bot detection strategy to unveil bots in the dataset at hand.

\subsubsection*{Summary: Temporal Dynamics.}
\begin{figure*}[t]
\includegraphics[width=\columnwidth]{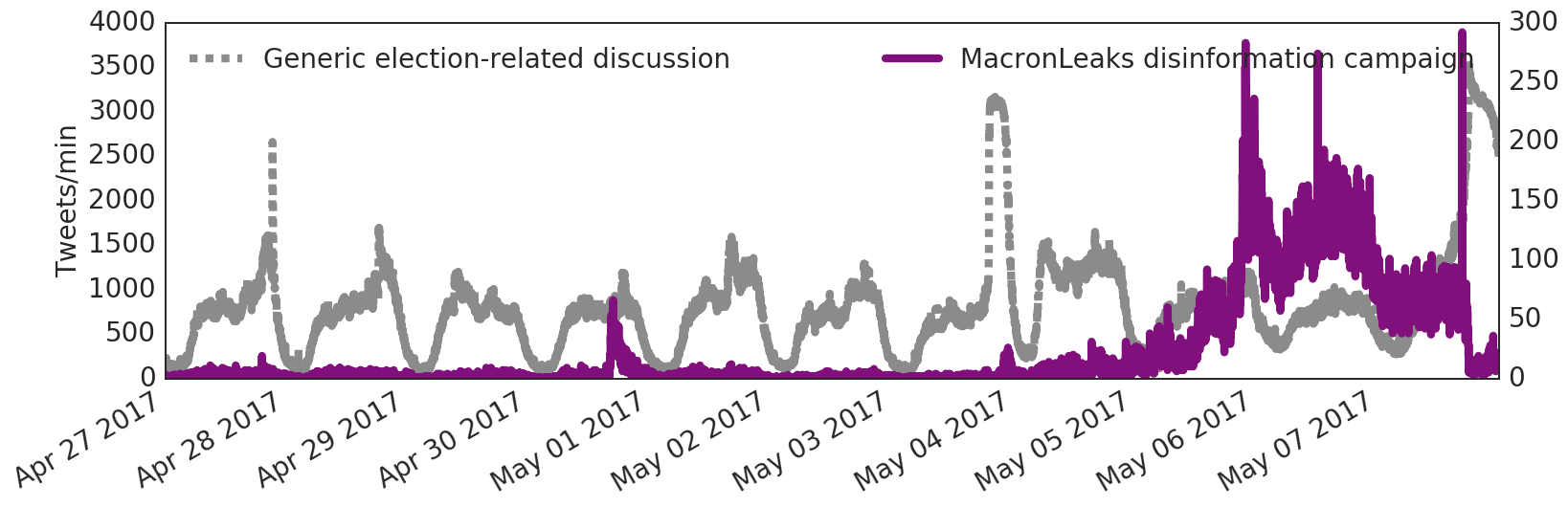}
\caption{Timeline of the volume of tweets generated every minute during our observation period (April 27 through May 7, 2017). The purple solid line (right axis) shows the volume associated with MacronLeaks, while the dashed grey line (left axis) shows the volume of generic election-related discussion. The presidential election occurred on May 7, 2017 (from~\cite{ferrara2017disinformation}).}
\label{fig:macron_timeline}
\end{figure*}

We started by exploring the timeline of the general election-related discussion on Twitter. The broader discussion that we collected concerns the two candidates, Marine Le Pen and Emmanuel Macron, and spans the period from April 27 to May 7, 2017, the Presidential Election Day, see Figure \ref{fig:macron_timeline}. Let us discuss first the dashed grey line (left axis): this shows the volume of generic election-related discussion. The discussion exhibits common circadian activity patterns and a slightly upwards trend in proximity to Election Day, and spikes in response to an off-line event, namely the televised political debate that saw Le Pen facing Macron. Otherwise, the number of tweets per minute averages between 300 and 1,500 during the day, and quickly approaches de facto zero overnight, consistently throughout the entire observation window. Figure~\ref{fig:macron_timeline} also illustrates with the purple solid line (right axis) the volume associated with MacronLeaks, the disinformation campaign that was orchestrated to smear Macron's reputation. The temporal pattern of this campaign is substantially different from the general conversation. First, the campaign is substantially silent for the entire period till early May. We can easily pinpoint the inception of the campaign on Twitter, which occurs in the afternoon of  April 30, 2017. After that, a surge in the volume of tweets, peaking at nearly 300 per minute, happens in the run up to Election Day, between May 5 and May 6, 2017. It is worth noting that such a peak is nearly comparable in scale to the volume of the regular discussion, suggesting that for a brief interval of time (roughly 48 hours) the MacronLeaks disinformation campaign acquired significant  attention~\cite{ferrara2017disinformation}.

\subsubsection*{Summary: Bot Dynamics.}

\begin{figure*}[t]
\includegraphics[width=\columnwidth]{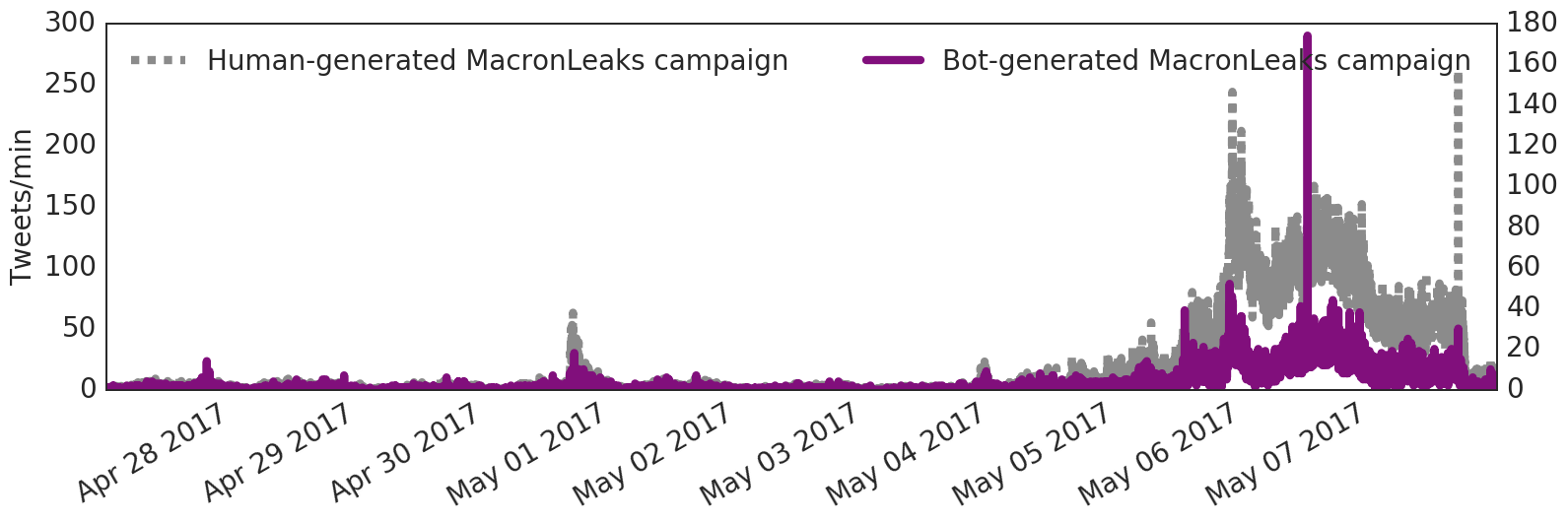}
\caption{Timeline of the volume of tweets generated every minute, respectively by human users (dashed grey line) and social bots (solid purple line), between April 27 and May 7, 2017, and related to MacronLeaks. Spikes in bot-generated content often slightly precedes spikes in human posts, suggesting that bots can trigger cascades of disinformation (from~\cite{ferrara2017disinformation}).}
\label{fig:macron_bots}
\end{figure*}

Like in the previous study, we here provide a  characterization of the Twitter activity, this time specifically related to MacronLeaks, for both bot and human accounts. In Figure~\ref{fig:macron_bots}, we show the timeline of the volume of tweets generated respectively by human users (dashed grey line) and bots (solid purple line), between  April 27 and May 7, 2017, and related to MacronLeaks. The amount of activity is substantially  close to zero until May 5, 2017, in line with the first coordination efforts as well as the information leaks spurred from other social platforms, as discussed in the paper~\cite{ferrara2017disinformation}. Spikes in bot-generated content often appear to slightly precede spikes in human posts, suggesting that bots can trigger cascades of disinformation~\cite{shao2018spread}. At peak, the volume of bot-generated tweets is comparable with the that of human-generated ones.
Further investigation revealed that the users who engaged with bots pushed the \textit{MacronLeaks} disinformation campaign were mostly foreigners with pre-existing interest in alt-right topics and alternative news media, rather than French users. Furthermore, we highlighted an anomalous account usage pattern where hundreds of bot accounts used in the 2017 French Presidential elections were also present in the 2016 US Presidential Election discussion, which suggested the possible existence of a black market for reusable political disinformation bots~\cite{ferrara2017disinformation}.

\subsubsection*{Summary: Sentiment Dynamics.}
\begin{figure}[h!]\centering
\includegraphics[width=.7\columnwidth]{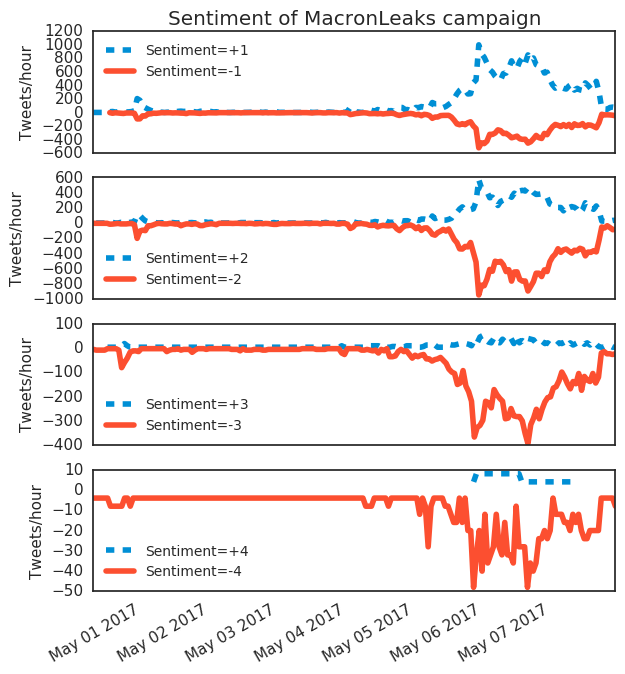}
\caption{Temporal distribution of sentiment disaggregated by sentiment intensity (hourly resolution). The sign on the y-axis captures the amount of tweets in the positive (resp., negative) sentiment dimension.}
\label{fig:macron_sentiment}
\end{figure}

Identically to the 2016 US Presidential Election study, we annotated all tweets in this corpus using \textit{SentiStrength}, and subsequently studied the evolution of the sentiment of tweets in the 2017 French Presidential Election discussion. Figure~\ref{fig:macron_sentiment} shows the temporal distribution of tweets' sentiment disaggregated by intensity: the four panels illustrate the overall timeline of the volume of tweets that exhibit positive and negative sentiment at the hourly resolution, for sentiment polarities ranging from 1 (lowest) to 4 (highest) in both positive and negative spectra. What appears evident is that, as  Election Day approaches, moderately and highly negative tweets (sentiment scores of -2, -3, and -4) significantly outnumber the moderately and highly positive tweets, at times by almost an order of magnitude. For example, between May 6 and 7, 2017, on average between 300 and 400 tweets with significant negative sentiment (sentiment scores of -3) were posted every hour, compared with an average of between 10 and 50 tweets with an equivalently positive sentiment (score scores of +3). Since the discussion during that period was significantly driven by bots, and bots focused against Macron, our analysis suggested that bots were pushing  negative campaigns against that candidate aimed at smearing his credibility and weakening his position in the eve of the May 7's election.

\subsection{2018 US Midterms}

The notorious investigation on Russian interference led by the US Senate Select Committee on Intelligence (SSCI) put social media service providers (SMSPs) at the center-stage of the public debate.
According to reports, SMPSs started to devote more efforts to ``sanitize'' their platforms, including ramping up the technological solutions to detect and fight abuse. Much attention has been devoted to identifying and suspending \textit{inauthentic activity}, a term that captures a variety of tools used to carry out manipulation, including bot and troll accounts.

Hence, it is natural to ask whether these countermeasures proved effective, or if otherwise the strategies and technologies bots typically used until 2017 evolved, and to what extent they successfully adapted to the changing social media defenses and thus escaped detection. We recently set to answer these questions: to this purpose, we monitored and investigated the online activity surrounding the 2018 US Midterm elections what were held on November 6, 2018. 

\subsubsection*{Data Collection.} 
We collected data for six weeks, from October 6, 2018 to November 19, 2018, i.e., one month prior and until two weeks after election day.
Tweets were collected using the Twitter Streaming API and following these keywords: \textit{2018midtermelections}, \textit{2018midterms}, \textit{elections}, \textit{midterm}, and \textit{midtermelections}.
Post-processing and cleaning procedures  are described in detail in our paper \cite{luceri2019red}: we retained only tweets in English, and manually removed tweets that were out of context, e.g., tweets related to other countries' elections (Cameroon, Congo, Biafra, Kenya, India, etc.) that were present in our initial corpus because they contained the same keywords we tracked. The final dataset contains 2.6M tweets, posted by nearly 1M users.

\subsubsection*{Bot Detection.}
Similarly to the 2016 US Presidential election study, since this study was a post-mortem (i.e., not in real time but after the events), we adopted Botometer to infer the bot scores of the users in our dataset. The only distinction worth mentioning is that we used the Botometer API version \textit{v3} that brings new features and a non-linear re-calibration of the model: in line with the associated study's recommendations \cite{yang2019arming}, we used a threshold of $0.3$ (which corresponds to a $0.5$ threshold from previous versions of Botometer) to separate bots from humans (note  that the results remain substantially unchanged if a higher threshold was used).
As a result, we obtained that 21.1\% of the accounts were categorized as bots, which were responsible for 30.6\% of the total tweets in our dataset. Manual validation procedures assessed the reasonable quality of these annotations. 
The resulting evidence suggests that bots were still present, and accounted for a significant amount of the tweets posted in the context of the political discourse revolving around the 2018 US Midterms. 

Interestingly, about 40 thousand accounts were already inactive at the time of our analysis, and thus we were not able to infer their bot scores using the Twitter API. We manually verified that 99.4\% of them were suspended by Twitter, corroborating the hypothesis that these were bots as well, and were suspended by Twitter in the time between the events and our post-mortem analysis, which was carried out in early 2019.

\subsubsection*{Political Leaning Inference.}
Next, we set to determine if bots exhibited a clear political leaning, and if they acted according to that preference. To label accounts as conservative or liberal, we used a label propagation approach that leveraged the political alignment of news sources whose URLs were posted by the accounts in the dataset. Lists of partisan media outlets were taken from third-party organizations, namely AllSides.Org and MediaBiasFactCheck.Com. 
The details of our label propagation algorithm are explained in our paper \cite{luceri2019red}. Ultimately, the procedure allowed us to reliably infer, with accuracy above 89\%, the political alignment of the majority of human and bot accounts in our corpus. These were factored into the subsequent analyses aimed at determining partisan strategies and narratives (see \cite{luceri2019red}).

\subsubsection*{Summary: Bot Activity and Strategies.}
Provided the evidence that bots were still present despite the efforts of the SMSPs to sanitize their platforms, we aimed at determining the degree to which they were embedded in the human ecosystem, specifically in the retweet network. This network is of central importance in our analysis, because it conveys information diffusion dynamics; many recent studies suggested a connection between bots and the spread of unverified and false information~\cite{shao2018spread,vosoughi2018spread}. It is therefore of paramount importance to determine if bots still played a role in the retweet network of election-related social media discourse as of 2018.

To this aim, we resorted to perform the $k$-core decomposition analysis. In social network theory, a $k$-core is a subgraph of a graph where all nodes have degree at least equal to $k$. The intuition is that, as $k$ grows, one is looking at  increasingly more highly-connected nodes' subgraphs. Evidence suggests that high $k$-cores are associated with nodes that are more embedded, thus influential, for the network under investigation \cite{wasserman1994social}. 

If bots were still influential in the 2018 US Midterm election discussion, our hypothesis is that we would find them in high concentration predominantly into high $k$ cores. This would be consistent with our findings related to the 2016 US Presidential Election discussion \cite{bessi2016social}.

\begin{figure}[t] 
\centering
  \includegraphics[width=.7\columnwidth]{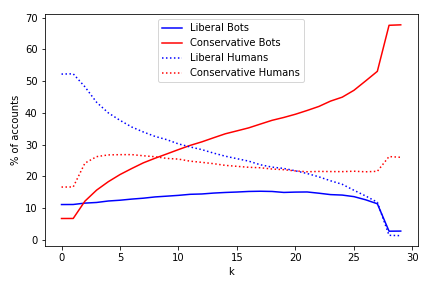}
  \caption{K-core decomposition: liberal vs. conservative bots and humans (from \cite{luceri2019red}).}
  \label{fig:core}
\end{figure}

Figure \ref{fig:core} corroborates our intuition. Specifically, we show the percentage of both conservative and liberal human and bot accounts as a function of varying $k$. Two patterns are worth discussing: first, as $k$ increases, the fraction of conservative bots grows, while the prevalence of liberal bots remains more or less constant; conversely, the prevalence of human accounts decreases, with growing $k$, more markedly for liberal users than conservative ones. We summarize these findings suggesting that conservative bots were situated in a premium position in the retweet network, and therefore may have affected information spread \cite{luceri2019red}.

\subsection{2016 vs 2018: A Comparative Bot Analysis}
Having identified and analyzed the activity of human and bot accounts in the context of the political discourse associated to US election events in both 2016 and 2018, it is natural to ask whether these studies involved a similar set of accounts. In other words, it is worth determining whether there exists a continuum of users that are active in both time periods under investigation. If this is the case, it would be interesting to study the users present in both periods, determine whether any of them are the bots under scrutiny in the previous studies, and ultimately understand if the strategies they may have exhibited evolved, possibly to escape detection or avoid further scrutiny of SMSPs.

\subsubsection*{Data Collection.} 
To answer the questions above, we isolated the users present in both the 2016 and 2018 datasets described above. This process yielded over 278 thousand accounts, active in both periods. Further processing and cleaning procedures, as detailed in our paper \cite{luceri2019evolution}, brought the dataset down to 245K users, accounting for over 8.3M tweets in 2016 and 660K in 2018. Botometer was used to determine the bot scores of these accounts. As a result, 12.6\% of  these accounts scored high in bot scores and were therefore classified as bots. We used this dataset to study the evolution of behavior of bots over the time period of study.

\subsubsection*{Summary: Bot Evolution Dynamics.}

One advantage of bots over humans is their scalability. Since bots are controlled by software rather than human users, as such they can work over the clock, they don't need to take rests and don't have the finite cognitive capacity and bandwidth that dictates how humans operate on social media \cite{pozzana2018measuring}. In principle, a bot could post continuously without any break, or at regular yet tight intervals of time. As a matter of fact, primitive bots used these simple strategies \cite{ratkiewicz2011detecting,metaxas2012social}. However, such obvious patterns are easy to spot automatically, hence not very effective. There is therefore a trade-off between realistic-looking activity and effectiveness. In other words, one can investigate the patterns of \textit{inter-event time} betweet a tweet post and its subsequent, and lay out the frequency distribution in an attempt to distill the difference between human and bot accounts' temporal dynamics. 

\begin{figure*}[t]
\includegraphics[width=\columnwidth]{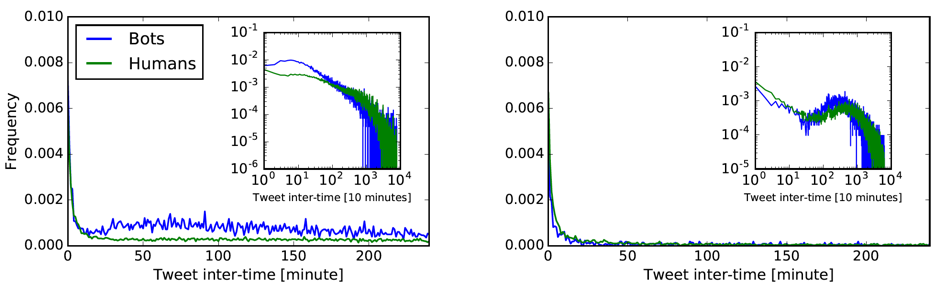}
\caption{Tweet inter-event time by bots and humans in 2016 (left) and 2018 (right). A clear distinction in temporal signature between bots and humans was evident in 2016, but vanished in 2018 (from~\cite{luceri2019evolution}).}
\label{fig:intertime}
\end{figure*}

Figure \ref{fig:intertime} illustrates the tweet inter-time distribution by bots and humans in 2016 (left) and 2018 (right). It is apparent that, while in 2016 bots exhibited a significantly different frequency distribution with respect to their human counterparts, in 2018 this distinction has vanished. In fact, statistical testing of distribution differences suggests that human and bot temporal signatures are indistinguishable in 2018. The discrepancy is particularly relevant in the time range between 10 minutes and three hours, consistent with other findings \cite{pozzana2018measuring}: in 2016, bots shared content at a higher rate with respect to human users.

Our work \cite{luceri2019evolution} corroborates the hypothesis that bots are continuously changing and evolving to escape detection. Further examples that we reported also illustrate other   patterns of behavior that have changed between 2016 and 2018: for instance, the sentiment that was expressed in favor or against political candidates in 2018 reflects significantly better what the human crowd is expressing. However, in 2016, bots' sentiment drastically diverged, in a manner easy to detect, from that of the human's counterpart, as we discussed earlier.

\section{Conclusions}

In this chapter, we set to discuss our latest results regarding the role of bots within  online political discourse in association with three major political events. 

First, we described the results of our analysis that unveiled a significant amount of bots distorting the online discussion in relation to the 2016 US Presidential election. We characterized the activities of such bots, and illustrated how they successfully fostered interactions by means of retweets at the same rate human users did. Other researchers suggested that this played a role in the spread of false news during that time frame \cite{shao2018spread,vosoughi2018spread}.

Second, we highlighted the role of bots in pushing a disinformation campaign, known as \textit{MacronLeaks}, in the run up to the 2017 French Presidential election. We demonstrated how it is possible to easily pinpoint the inception of this disinformation campaign on Twitter, and we illustrated how its popularity peak was comparable with that of regular political discussion. We also hypothesized that this disinformation campaign did not have a major success in part because it was tailored around the information needs and usage patterns of the American alt-right community rather than French-speaking audience. Moreover, we found that several hundreds of bot accounts were re-purposed from the 2016 US Election. Ultimately, we suggested the possibility that a black market for reusable political bots may exist \cite{ferrara2017disinformation}. 

Third, we studied the 2018 US Midterms, to investigate if bots were still present and active. Our analysis illustrated that not only bots were almost as prevalent as in the two other events, but also that conservative bots played a central role in the highly-connected core of the retweet network. These findings further motivated a comparative analysis contrasting the activity of bots and humans in 2016 and 2018. Our study highlighted that a core of over 245K users, of which 12.1\% were bots, was active in both events. Our results suggest that  bots may have  evolved to better mimic human temporal patterns of activity.

With the increasing sophistication of Artificial Intelligence, the ability of bots to mimic human behavior to escape detection is greatly enhanced. This poses challenges for the research community, specifically in the space of bot detection. Whether it is possible to win this arms race is yet to be determined: any party with significant resources can deploy state of the art technologies to enact influence operations and other forms of manipulation of  public opinion. 

The availability of powerful neural language models lowers the bar to adopt techniques that allow to build credible bots. For example, it may be already in principle possible to automatize almost completely the generation of genuine-looking text. This may be used to push  particular narratives, to  artificially build  traction for political arguments that may  otherwise have little or no human organic support. 

Ultimately, the evidence that our studies, and the work of many other researchers in this field, have brought strongly suggest that more policy and regulations may be warranted, and that technological solutions alone may not be sufficient to tackle the issues of bot interference in political discourse. 

\section*{Acknowledgements}
The author is grateful to his collaborators and coauthors on the topics covered in  this paper, in particular Adam Badawy, Alessandro Bessi, Ashok Deb, and Luca Luceri, who  contributed significantly  to three papers widely discussed in this chapter \cite{bessi2016social,luceri2019red,luceri2019evolution}.

\bibliographystyle{abbrv}
\bibliography{bots}
\end{document}